\RequirePackage{lineno}
\documentclass{nature}
\usepackage{xcolor}
\usepackage{amsmath}
\usepackage{caption}
\usepackage{graphicx}
\usepackage{tabularx}
\usepackage{makecell}
\makeatletter
\let\saved@includegraphics\includegraphics
\AtBeginDocument{\let\includegraphics\saved@includegraphics}
\renewenvironment*{figure}{\@float{figure}}{\end@float}
\makeatother
\usepackage{graphicx}
\makeatletter
\let\saved@includegraphics\includegraphics
\AtBeginDocument{\let\includegraphics\saved@includegraphics}

\makeatother

\title{Observation of Floquet-induced gap in graphene}

\author{Fei Wang$^{1,2,\ast}$, Xuanxi Cai$^{1,2,\ast}$, Xiao Tang$^{1,2}$, Jinxi Lu$^{1,2}$, Wanying Chen$^{1,2}$, Tianshuang Sheng$^{1,2}$, Runfa Feng$^{1,2}$, Haoyuan Zhong$^{1,2}$, Hongyun Zhang$^{1,2}$, Pu Yu$^{1,2,3}$ \& Shuyun Zhou$^{1,2,3,\dagger}$}

\begin{document}
\maketitle

\begin{affiliations}

\item Department of Physics, Tsinghua University, Beijing 100084, People's Republic of China
\item State Key Laboratory of Low-Dimensional Quantum Physics, Tsinghua University, Beijing 100084, People's Republic of China
\item Frontier Science Center for Quantum Information, Beijing 100084, People's Republic of China

* These authors contributed equally to this work\\
$\dagger$ Correspondence and request for materials should be sent to syzhou@mail.tsinghua.edu.cn

\end{affiliations}

\begin{abstract}

Floquet engineering provides a powerful pathway for creating non-equilibrium phases of matter with tailored electronic structures and properties through time- periodic driving. As the original theoretical prototype, graphene established the framework in which the Floquet topological insulator with light-induced anomalous Hall effect was proposed. However, the defining spectroscopic signature of Floquet engineering in graphene--light-induced hybridization (avoided-crossing) gap at Floquet band crossings, has remained experimentally elusive. Here, we report direct observation of Floquet-induced hybridization gap in monolayer graphene under resonant driving by a strong light field. Time- and angle-resolved photoemission spectroscopy reveals gap opening at Floquet band crossings, accompanied by coherent Floquet sidebands. The gap exhibits pronounced momentum anisotropy, featuring two Dirac nodes protected by the spatiotemporal symmetry and tunable by light polarization. These results provide long-sought experimental demonstration of Floquet band engineering in graphene, opening up opportunities for light-field engineered quantum phases in graphene and related materials. 

\end{abstract}

Energy gaps are fundamental to quantum phases of matter. In the canonical solid-state physics model\cite{AshcroftMermin}, a spatially-periodic potential folds the free-electron dispersion (Fig.~1a) into momentum-periodic Bloch bands (Fig.~1b), yet such folding alone does not fundamentally change electronic properties. The decisive transformation occurs when these bands hybridize through interaction with the crystal potential, opening energy gaps at the Brillouin zone (BZ) boundary (Fig.~1c) which electronically distinguish semiconductors and insulators from metals. 
Time-periodic driving fields extend this conceptual framework to the nonequilibrium regime. 
By dressing Bloch electrons in the time domain, the driving field imposes an additional periodicity in energy\cite{Shirley1965Floquet}, forming Floquet-Bloch bands (Fig.~1d,e). Analogous to Bloch bands, the pivotal step is the opening of a Floquet-induced hybridization (avoided-crossing) gap at the Floquet BZ boundary (Fig.~1f), arising from Floquet band mixing through coherent light-matter interactions. Such light-induced gap is a hallmark of light-field tailored electronic states, the foundation of Floquet engineering. 

Graphene, a two-dimensional Dirac material with linear dispersion, provides an ideal platform to explore this paradigm. A Floquet-induced gap was first predicted\cite{EfetovPRB2008,NaumisPRB2008,oka2009prb}, together with a light-induced anomalous Hall effect\cite{oka2009prb}, the optical analogue of the Haldane model\cite{haldane1988model}. These seminal predictions established graphene as a model system for Floquet engineering, further inspiring extensive research on light-field control  across diverse platforms, from solid-state materials\cite{HsiehNM2017,oka2019floquet,SentefRMP2021,lindner2011NP,ZhouNRP2021} to photonic lattices\cite{OzawaRMP2019} and ultracold atoms\cite{EckardtRMP2017}. While Floquet engineering has been demonstrated in highly tunable synthetic systems such as photonic lattices\cite{Szameit2013photonic} and ultracold atoms\cite{Esslinger2014coldatom}, its realization in graphene has remained exceptionally challenging due to dissipation and interactions in solid-state materials\cite{Cavalleri2013NM,gierz2015tracking,Damascelli2019science,GierzNanoLett2021,ZhouNSR2021,Gil2019prb,Mitra2014prb,sato2019microscopic}.

Recent studies have revealed transport signatures of light-induced anomalous Hall effect in graphene\cite{Cavalleri2020np} and Floquet-Bloch states manifested through Floquet-Andreev states\cite{lee2022nat} and Floquet-Volkov interference\cite{gedik2025graphene,stefan2025graphene,DaniNPNV2025}, however, direct spectroscopic evidence of Floquet-induced gap, the defining signature of Floquet engineering, has remained elusive for a decade. 
Considering that such gap formation has been reported in topological insulator\cite{Gedik2013} and black phosphorus\cite{zhou2023pseudospin}, its absence in graphene, the simplest and theoretically best-understood model system\cite{EfetovPRB2008,NaumisPRB2008,oka2009prb,NaumisPhilosMag2010,Demler2011prb,AuerbachGRPRL2011,AlexandrePRB2011,TorresAPL2011,wu2011prb,PlateroPRB2013,Balseiro2014PRB,SavelevTimeCrysPRB2014,Devereaux2015NC,MengSGPRB2019,LudwigPRR2022}, is particularly striking and raises a fundamental question: can Floquet-engineered phases in graphene be realized under realistic experimental conditions? More broadly, can the elegant framework of Floquet theory faithfully describe real materials where dissipation and many-body interactions are difficult to avoid? Observing the Floquet-induced gap would therefore provide long-sought proof of principle for Floquet band engineering in this model system.

\begin{figure*}[htbp]
	\centering
	\includegraphics[width=16.8cm]{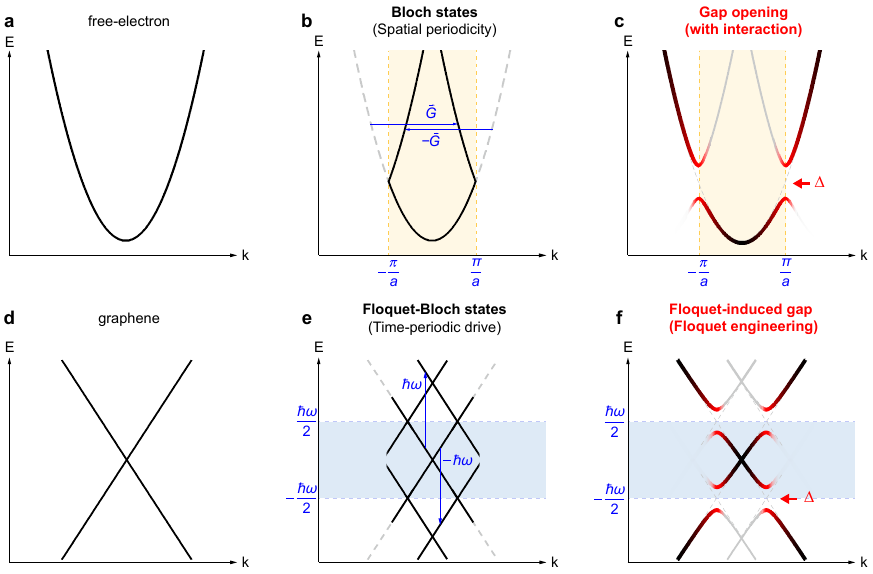}
	\caption*{{\bf Fig.~1 ${\mid}$  Schematic illustration of Bloch states, Floquet-Bloch states and gap opening.}
	 {\bf a}, Dispersion of free electrons.  
   {\bf b}, Folding of the electronic bands by the reciprocal lattice vector $\vec{G}$ under a spatially-periodic potential with period $a$ being the lattice constant.
   {\bf c}, Gap opening at the Brillouin zone (BZ) boundary due to interaction of electrons with the realistic spatially-periodic potential.  
	 {\bf d}, Dispersion of graphene. 
   {\bf e}, Floquet sidebands under a time-periodic drive, where $\hbar\omega$ is the driving photon energy.  
   {\bf f}, Floquet-induced gap at the Floquet BZ boundary through the strong light-matter interaction.  The yellow and blue shaded areas mark the first BZ in the momentum and energy space respectively.
	}
\label{Fig1}
\end{figure*}

Here we report the successful observation of the Floquet-induced hybridization gap, providing hallmark evidence of Floquet engineering in graphene, by synergistically integrating experimental advances on sample quality, driving field as well as instrumental resolutions into time- and angle-resolved photoemission spectroscopy (TrARPES) measurements. The gap opening is observed at Floquet band crossings, accompanied by Floquet sidebands. The gap exhibits strong momentum anisotropy, vanishing along the light polarization direction and maximized in the perpendicular direction, directly manifesting the spatiotemporal symmetry of the driven electronic system. These findings provide definitive experimental evidence of Floquet band engineering in graphene, establishing it as a  benchmark platform for realizing light-field driven quantum phases.

\section*{Experimental observation of light-induced gap in graphene}

We used an epitaxial monolayer graphene grown on SiC substrate\cite{syzhou2007NM} in this study. Figure~2a shows dispersion image measured through the K point perpendicular to the $\Gamma$-K direction, as indicated by the red line in the inset. The sample is electron doped with the Dirac point below the Fermi energy ($E_F$), and the dispersion is characterized by a conical dispersion with a gap at the Dirac point induced by graphene-substrate interaction\cite{syzhou2007NM,SonYWPRL08} and mid-gap states, as schematically illustrated in Fig.~2b. 

\begin{figure*}[htbp]
	\centering
	\includegraphics[width=16.8cm]{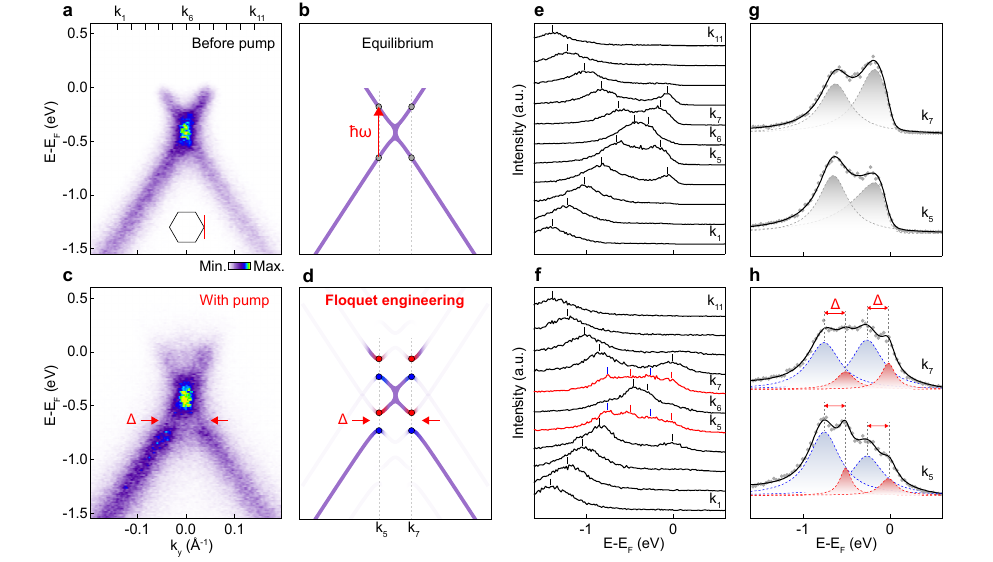}
	\caption*{{\bf Fig.~2 ${\mid}$  Experimental observation of light-induced gap in monolayer graphene.}
	 {\bf a,b}, Dispersion image measured before pumping ({\bf a}) and schematic dispersion ({\bf b}). The measurement direction is indicated by the red line in the inset (perpendicular to the $\Gamma$-K direction). The gray dots in ({\bf b}) mark the resonance points upon driving at $\hbar\omega$ = 490 meV.
	 {\bf c,d}, Dispersion image measured upon pumping at $\hbar\omega$ = 490 meV with pump fluence of 4.1 mJ/cm$^2$ ({\bf c}) and schematic dispersion ({\bf d}). The pump polarization is along the $\Gamma$-K direction. The Floquet-induced gap is labelled by $\Delta$. 
	 {\bf e,f}, EDCs at resonance points ($k_5$ and $k_7$) indicated in ({\bf d}). Tick marks indicate peak positions in the EDCs.
	 {\bf g,h}, Zoom-in EDCs at resonance points with fitting peaks appended.
}\label{Fig2}
\end{figure*}

We applied resonant driving with mid-infrared (MIR) wavelength of $\lambda$ = 2.53 $\mu$m (photon energy of $\hbar\omega$ = 490 meV). Due to the electron doping in the epitaxial graphene, this photon energy is insufficient to excite carriers above the Fermi energy (Fig.~2b) thereby suppressing photo-excited carriers, and the pump mainly acts as a time-periodic driving field. Under such light field, the first-order sideband ($n$ = $\mp$1) of the upper/lower Dirac cone overlaps with the lower/upper Dirac cone at the Floquet crossing points, $E_k^{VB}= E_k^{CB}-\hbar\omega$. The energies of these resonance points (marked by dots in Fig.~2b) define the Floquet Brillouin zone (Floquet BZ) boundaries in the energy domain, where Floquet-induced hybridization gap is expected.  The high-quality graphene sample with reduced scattering allows it to survive at a large pump fluence of $F$ = 4.1 mJ/cm$^2$ without significant pump-induced spectral broadening (Extended Data Fig.~1). This together with other experimental advances (see a comparison of experimental parameters with those in the literature in Extended Data Table~1) is critical for the observation of Floquet-induced gap in this model system.

Figure~2c shows dispersion image measured at delay time $\Delta t$ = 0 upon pumping with light polarization along the $\Gamma$-K direction. A striking modification of the transient electronic structure is clearly observed. In particular, there is a strong suppression of intensity (marked by red arrows in Fig.~2c) around the resonance points, indicating gap opening arising from Floquet band mixing, as schematically illustrated in Fig.~2d. 
The light-induced hybridization gap is further supported by energy distribution curve (EDC) analysis shown in Fig.~2e-h. The conical dispersion before pumping is revealed by following the positions of two dispersing peaks in Fig.~2e. Upon pumping, EDCs away from the resonance points (see black curves in Fig.~2f) show similar peaks as the equilibrium state, while EDCs at the resonance points ($k_5$ and $k_7$, see red curves in Fig.~2f) show clearly multiple-peak structure. Figure~2g,h shows zoom-in EDCs at these momentum points together with fitting results. The two peaks in each EDC (Fig.~2g), corresponding to electronic states in the valence band (VB) and conduction band (CB) in the equilibrium state, clearly split into four peaks upon pumping (Fig.~2h), signaling a gap opening at the resonance points both for the VB and CB. From the energy splitting of the fitting peaks, the light-induced hybridization gap is extracted to be $\Delta$ = 241 $\pm$ 18 meV, much larger than our experimental energy resolution\cite{syzhou2024hhg}, allowing the light-induced hybridization gap to be resolved unambiguously for the first time.

\section*{Evidence for Floquet origin of the gap}

\begin{figure*}[htbp]
	\centering
	\includegraphics[width=16.8cm]{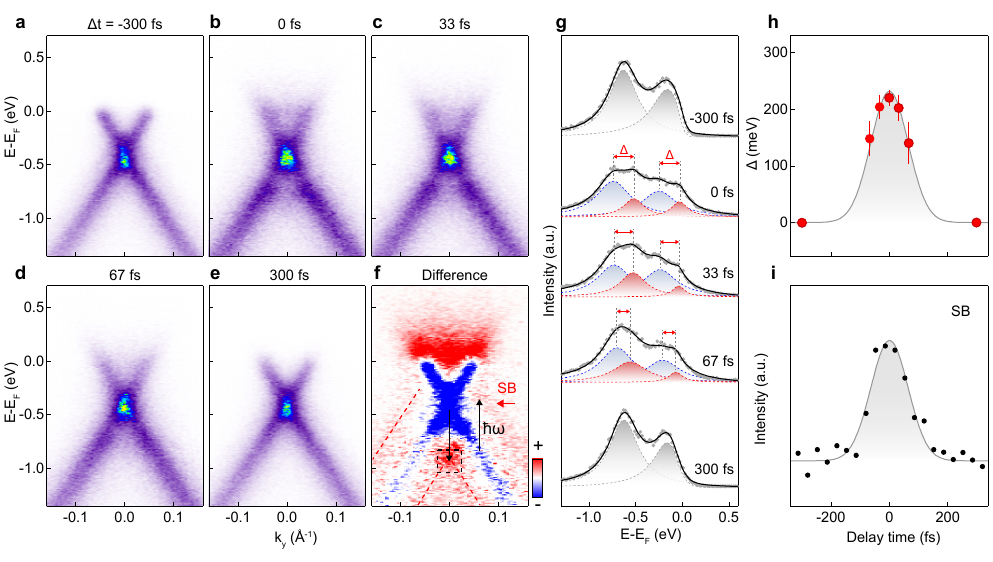}
	\caption*{{\bf Fig.~3 ${\mid}$ Evidence of Floquet-induced gap from time-dependent measurements.}
     {\bf a-e}, Dispersion images measured at different delay times upon driving at $\hbar\omega$ = 490 meV with pump fluence of 3.8 mJ/cm$^2$.
	 {\bf f}, Differential image obtained by subtracting data measured at $\Delta$t = -300 fs ({\bf a}) from 0 fs ({\bf b}). Red arrow and black dashed box indicate the light-induced sideband (SB).
     {\bf g}, EDCs at resonance points measured at different delay times with fitting peaks appended. 
	 {\bf h},  Extracted Floquet-induced gap at different delay times. The gray curve is the same as that in ({\bf i}).
	 {\bf i}, Integrated intensity of the light-induced sideband over the box in ({\bf f}) as a function of delay time. The gray curve is the fitting curve of the sideband intensity.
	}\label{Fig3}
\end{figure*}

\section*{Evidence for Floquet origin of the gap}

The Floquet origin of this gap is supported by time-dependent measurements, which show that the gap opening occurs only when the driving field is on, and is accompanied by coherent Floquet sidebands. Figure~3a-e shows snapshots of dispersion images measured at different delay times. The gap is clearly resolved near time zero when the pump field is on, and becomes vanishing at $\pm$300 fs when the pump does not overlap with the probe. The sidebands, which are weak replicas of Dirac cone shifted by the pump photon energy (indicated by the red arrow and black box in the differential image in Fig.~3f), are also observed near time zero. We note that in our experimental geometry (see Extended Data Fig.~2), the pump is incident near normal to the sample surface, resulting in a driving field predominantly within the sample plane. The negligible out-of-plane field component suppresses the Volkov states, that is, light-field dressed photoemission final states\cite{Gedik2016}, leading to weaker sideband intensity compared to recent reports\cite{gedik2025graphene,stefan2025graphene}; meanwhile, the stronger in-plane field component also has the advantage of generating a larger Floquet-induced gap.

We further analyze the temporal evolution of both the extracted gap size and the sideband intensity. The extracted gap, obtained by fitting EDCs at the resonance points in Fig.~3g, is plotted as a function of delay time in Fig.~3h. Comparison with the integrated sideband intensity in Fig.~3i shows that both the gap opening and the sidebands are observed within a temporal window of 150 fs, which corresponds to the instrumental time resolution set by the pump-probe overlap. This suggests that they are both driven by the coherent light fields, confirming their Floquet origin from the time domain perspective.

In addition to the temporal evolution, pump fluence and photon energy dependent measurements also support the Floquet origin of the observed gap. Extended Data Fig.~3 shows that the light-induced gap $\Delta$ scales with the pump fluence $F$ and the light field $E$ by $\Delta \propto \sqrt{F} \propto E$, consistent with the expected scaling in the Floquet engineering framework\cite{Galitski2013PRB,Gedik2016,zhou2023pseudospin}. Furthermore, the Floquet-induced gap is also observed when driving at 600 meV photon energy, at momentum positions farther away from the K point compared to driving at 490 meV (Extended Data Fig.~4), which is also consistent with the Floquet origin of this gap.

\section*{Anisotropic gap with spatiotemporal symmetry protected Dirac nodes}

The Floquet-induced hybridization gap exhibits a pronounced anisotropy in the momentum space. Figure~4a-c shows representative dispersion images measured at selected azimuthal angles $\varphi$ through the K point, keeping the pump polarization along the $k_x$ direction (Fig.~4d). A gap opening is clearly observed when the electron momentum is perpendicular to the light field ($\varphi = 90^\circ$; Fig.~4a), which decreases and eventually vanishes when the electron momentum aligns with the light field ($\varphi = 0^\circ$; Fig.~4c). The extracted gap at different azimuthal angles (Fig.~4f) shows that the gap is maximized at $\varphi = 90^\circ$ and $270^\circ$, and vanishes at $\varphi = 0^\circ$ and $180^\circ$, thereby creating two Dirac nodes along the pump field direction as schematically illustrated in Fig.~4g.
The absence of Floquet-induced gap for a pump polarized along the $k_y$ direction (parallel to the electron momentum, see Extended Data Fig.~5) further confirms that the gap node emerges when the driving field is parallel to the electron momentum. 

\begin{figure*}[htbp]
	\centering	
	\includegraphics[width=16.8cm]{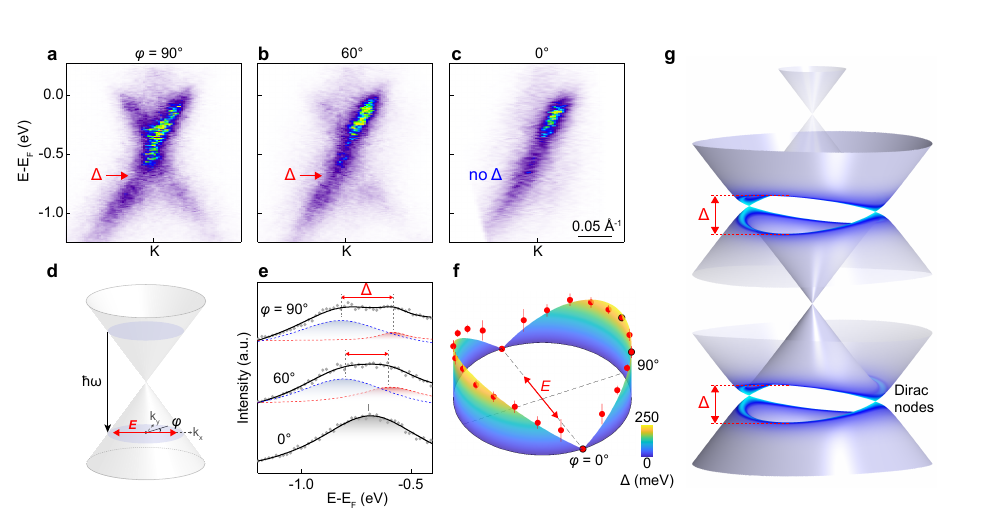}
  \caption*{{\bf Fig.~4 ${\mid}$ Momentum anisotropy of the Floquet-induced gap and the emergence of spatiotemporal symmetry protected Dirac nodes.}
   {\bf a-c}, Dispersion images measured at $\Delta t$ = 0 at azimuthal angle $\phi$ = 90$^\circ$, 60$^\circ$ and 0$^\circ$ upon pumping at $\hbar\omega$ = 490 meV with pump fluence of 4.1 mJ/cm$^2$.
   {\bf d}, Schematic illustration for the resonance rings in the VB and CB. The red arrow indicates the pump light field direction.
   {\bf e}, Zoom-in EDCs at the resonance points, together with fitting curves to extract the Floquet-induced gap for data shown in ({\bf a-c}).
   {\bf f}, Schematic illustration of the anisotropic gap opening along the resonance rings, and the emergence of two spatiotemporal symmetry protected Dirac nodes.
    }\label{Fig4}
\end{figure*}

The anisotropic Floquet-induced hybridization gap originates from the interplay between the unique pseudospin texture of graphene and the spatiotemporal symmetry of the driven system\cite{EfetovPRB2008,oka2009prb,wu2011prb,Galitski2013PRB}. In graphene\cite{NetoRMP09}, the Dirac Hamiltonian $H_0 = v_F\boldsymbol{\sigma}\cdot\mathbf{k}$ locks the pseudospin $\boldsymbol{\sigma}$ to the electron momentum $\mathbf{k}$. Under a linearly polarized drive, the perturbation $V(t) = \frac{e}{c} v_F\boldsymbol{\sigma}\cdot\mathbf{A}(t)$ acts as a time-dependent pseudomagnetic field, where $v_F$ is the Fermi velocity and $\mathbf{E}(t) = -\partial \mathbf{A}(t)/\partial t$ is the driving light field. For electron momentum perpendicular to the light field ($\mathbf{k}\perp\mathbf{E}$), the driving field introduces a transverse pseudospin component, resulting in $[H_0, V(t)] \neq 0$. This breaks the spatiotemporal symmetry and opens a hybridization gap at Floquet band crossings. In contrast, for $\mathbf{k}\parallel\mathbf{E}$, the perturbation $V(t)$ commutes with the Dirac Hamiltonian ($[H_0, V(t)] = 0$), namely, preserving the spatiotemporal symmetry. Such spatiotemporal symmetry protects the gapless Dirac crossings, allowing dynamic control of the momentum positions of the emergent Dirac nodes by rotating the pump polarization.

\section*{Discussion and outlook}

Our work not only resolves the long-standing challenge of realizing Floquet engineering in graphene but also establishes a few guiding principles for its experimental realization in quantum materials, which requires a critical balance between achieving strong light-matter coupling while preserving quantum coherence in dissipative solid-state environments. Here we summarize key experimental advances that are critical for the realization of Floquet engineering: (1) high-quality sample with reduced scattering to sustain high pump fluence; (2) a strong Floquet interaction strength $\beta\propto \dfrac{E}{\omega^2}$ achieved through resonant driving with field strength $E>10^8$ V/m and low driving photon energy $\hbar\omega$; (3)  suppression of photo-excited carriers by using a low pump photon energy which is insufficient for direct photo-excitation; (4) ultrashort probe pulse ($<$ 100 fs) to capture Floquet-induced effects before electron-phonon and other scatterings dominate at longer delay time\cite{Cavalleri2013NM} ; (5) high energy resolution enabled by optimized high-harmonic generation light source\cite{syzhou2024hhg} to resolve subtle gap features.

Our work provides definitive experimental evidence that time-periodic light field can reshape the low-energy Dirac band structure of graphene, an essential prerequisite for achieving light-induced topological phases, an important long-term and landmark research goal for this field. In epitaxial graphene, however, the equilibrium gap opening at the Dirac point\cite{syzhou2007NM,SonYWPRL08} together with the presence of mid-gap states hinders the clear identification of light-induced topological gap under circularly polarized driving. Future experiments should therefore employ intrinsically gapless monolayer graphene, such as exfoliated graphene flakes. Yet, given the typical micrometer-scale size of such samples, it is crucial to use HHG probe with tightly focused beam spot or to increase the sample size. It is also important to further enhance light-matter coupling strength, since the gap at the Dirac point results from a higher-order (and intrinsically weaker) interaction. Our findings thus establish a foundation and delineate clear pathways toward realizing Floquet topological insulators\cite{oka2009prb,Balseiro2014PRB}. Furthermore, our work paves the way for exploring Floquet time crystals\cite{SavelevTimeCrysPRB2014}, and moir\'e-Floquet engineering in van der Waals heterostructures\cite{SentefMFPRR2019,moire2021low}, where the interplay between moir\'e potentials and optical driving may give rise to emergent light-induced correlated phenomena.

\begin{addendum}
  \item [Acknowledgement] 
  This work is supported by the National Natural Science Foundation of China (Grant No.~12421004, 12234011), Tsinghua University Initiative Scientific Research Program (Grant No.~20251080106), National Key R\&D Program of China (Grant No.~2021YFA1400100), the National Natural Science Foundation of China (Grant No.~12327805, 52388201), and New Cornerstone Science Foundation through the XPLORER PRIZE.
 
  \item[Author Contributions] 
  S.Z. conceived the research project. F.W. and X.C. performed TrARPES measurements and analyzed the data. R.F., X.T. and W. C. prepared the samples. X.C. and Haoyuan Z. developed and optimized the HHG light source. F. W., X. C., J. L., T. S., Haoyuan Z., Hongyun Z. and P.Y. discussed the results. F.W., X.C. and S.Z. wrote the manuscript, and all authors commented on the manuscript.

  \item[Competing Interests] 
  The authors declare that they have no competing financial interests.
\end{addendum}

\bibliography{reference}

\begin{methods}

\subsection{Experimental setup.}
TrARPES measurements were performed in the home laboratory at Tsinghua University with a regenerative amplifier laser with a center wavelength of 800 nm (1.55 eV), a pulse energy of 1.3 mJ and a pulse duration of 35 fs, running at a repetition rate of 10 kHz. The fundamental beam was split into the probe and pump branches. The probe beam with photon energy of 21.7 eV was generated via high harmonic generation (HHG) approach, by focusing the second harmonic (3.1 eV) of the fundamental beam into an Argon-filled gas cell\cite{syzhou2024hhg}. The 21.7 eV pulse (7$^{th}$ harmonic of 3.1 eV) with pulse duration of 66 fs was isolated from other harmonics by passing through an aluminum foil and a tin foil. The selected 21.7 eV beam was subsequently focused onto the sample by a gold-coated toroidal mirror. The probe beam was $s$-polarized, which is almost aligned with $k_y$ direction of the graphene sample. The short probe pulse of 66 fs allows to capture Floquet-induced gap before electron-phonon and other scatterings dominate\cite{Cavalleri2013NM}, which typically occur on a longer timescale of a few hundred fs.
The pump beam with photon energy of 490 meV or 600 meV was generated by an optical parametric amplifier delivering pulses with durations of 135 fs and 113 fs, respectively. The pump beam was incident at approximately 5$^{\circ}$ with respect to the sample normal. The samples were measured at a temperature of 80 K in an ultra-high vacuum chamber with base pressure better than 2 $\times$ 10$^{-10}$ Torr. 

\subsection{Sample preparation.}
The epitaxial graphene was grown by thermal decomposition of SiC\cite{xue2013graphene}. Nitrogen-doped SiC (0001) substrate was cleaned with alcohol before transferring into ultrahigh vacuum chamber, and degassed at 650$^\circ$C for 3 hours by direct current heating. The graphitization was conducted by cycles of flash annealing from 650$^\circ$C to 1270$^\circ$C. Before TrARPES measurement, the graphene sample was annealed at 650$^\circ$C for 24 hours in an ultrahigh vacuum chamber. The Fermi energy is calibrated by fitting the momentum-integrated intensity curve near $E_F$ by a Fermi-Dirac distribution (Extended Data Fig.~1). Upon pumping, the energy distribution curve away from the gap region shows no significant broadening at different delay times (see Supplementary Information Fig.~S1).

\subsection{Extraction of the Floquet-induced gaps.}
To quantify the Floquet-induced hybridization gap at the resonance points, each EDC in Fig.~2 and Fig.~3 was obtained by integrating over a momentum window of 0.0108 $\mathrm{\AA}^{-1}$ and fitted with four Lorentzian peaks (corresponding to the splitting peaks in both the VB and CB) on a linear background. The gap size is defined by the energy separation between the two splitting peaks in VB/CB. The small fitting residuals (Extended Data Fig.~6) indicate the high-quality of the fitting. The uncertainty of the extracted gap is calculated as $\Delta = \sqrt{\Delta _{fit}^2+\Delta _{sys}^2}$, where $\Delta _{fit}$ is statistical fitting error and $\Delta _{sys}$ is the estimated systematic contribution from the choice of momentum integration window. When choosing different momentum integration windows, the extracted hybridization gap shows a variation of $\Delta _{sys}<10$ meV (Extended Data Fig.~7). In addition to fitting, a direct comparison between TrARPES data and the simulated spectrum (Extended Data Fig.~8) also supports that such gap is well resolvable in our experiments.

\subsection{Calculated Floquet-induced gap.}

The pump beam is incident on the sample surface at a near-normal angle of $\theta = 5^\circ$, as illustrated in Extended Data Fig.~2. The typical pump fluence of $4.1~\mathrm{mJ/cm^2}$ on the sample was calculated by using the formula:
\begin{equation}
F = \frac{P\times t}{f \times \pi \cdot \frac{D_1}{2} \cdot \frac{D_2}{2}},
\end{equation}
where $P$ is the average power of the pump beam measured outside the ARPES chamber, $t$ = 0.7 is the transmittance of the ZnSe window on the ARPES chamber, $f$ is the repetition rate of the laser, and $D_1$ and $D_2$ are the major and minor axes of the elliptical pump beam profile on the sample inside the ultrahigh vacuum chamber.

Using this fluence value, the electric field strength for the \textit{p}-polarized pump is $E = 4.8 \times 10^8~\mathrm{V/m}$, derived from the expression:
\begin{equation}
E = \sqrt{2 \cdot \frac{F}{\tau} \sqrt{\frac{\mu_0}{\epsilon_0}}},
\end{equation}
where $\tau = 135~\mathrm{fs}$ is the estimated pump pulse duration, $\mu_0$ and $\epsilon_0$ are the vacuum permeability and vacuum permittivity, respectively.

To determine the electric field parallel to the sample surface, Fresnel reflection needs to be considered. Based on Ref.~\cite{gr_refractive_index_2023}, the refractive index at the pump photon energy of $490~\mathrm{meV}$ is n = 4.32. Using Fresnel equations at an incidence angle of $5^\circ$, we obtain a transmission coefficient of $t_s$ = 0.38 for the electric field component parallel to the surface. This yields an in-plane electric field $E_i = 1.8 \times 10^8~\mathrm{V/m}$.

The dimensionless Floquet coupling strength is given by $\beta = \dfrac{e v_f E_i}{\hbar \omega^2}$ \cite{wu2011prb}, where $v_f = 1 \times 10^6~\mathrm{m/s}$ is the Fermi velocity of graphene. Substituting the experimental values yields $\beta = 0.49$. The corresponding Floquet-induced band gap is therefore estimated by
\begin{equation}
\Delta = \beta \cdot \hbar \omega = 240~\mathrm{meV}.
\end{equation}

\subsection{Anisotropic Floquet-induced hybridization gap from Floquet calculation.}

The Floquet band engineering of graphene and anisotropic Floquet-induced gap in Fig.~1c and Fig.~4g are obtained through Floquet calculation\cite{Galitski2013PRB,Balseiro2014PRB}. Only the first-order Floquet sidebands ($n=\pm 1$) are taken into account. Under \textit{p-pol.} pump along the $k_x$ direction, the effective Hamiltonian can be written as 
\begin{equation*}
	H_{\mathrm{Floquet}}^{{\mathrm{eff}}} = \left( {\begin{array}{*{20}{c}}
{\hbar{v_F}{\boldsymbol{\sigma }} \cdot {\bf{k}} + \hbar \omega {I}}&{\frac{V}{2}{\sigma _x}}&0\\
{\frac{V}{2}{\sigma _x}}&{\hbar{v_F}{\boldsymbol{\sigma }} \cdot {\bf{k}}}&{\frac{V}{2}{\sigma _x}}\\
0&{\frac{V}{2}{\sigma _x}}&{\hbar{v_F}{\boldsymbol{\sigma }} \cdot {\bf{k}} - \hbar \omega {I}}
\end{array}} \right)	
\end{equation*}
where $v_F$ is the Fermi velocity, $\bf{k}$ is the momentum measured from graphene K point, $\hbar \omega$ is the pump photon energy, and $V$ describes the Floquet coupling strength, $\boldsymbol{\sigma} = \sigma_x \hat{e}_x + \sigma_y \hat{e}_y$ is the Pauli  matrix, and $\boldsymbol{I}$ is the $2\times2$ unit matrix. The electronic structure of graphene upon pumping is obtained by solving the Floquet equation.

\end{methods}

\begin{addendum}
	\item[Data availability]
	All data supporting the results of this study are available within the article. Additional data are available from the corresponding author upon request.

  \item[Correspondence and requests for materials] should be addressed to Shuyun Zhou (syzhou@mail.tsinghua.edu.cn).

\end{addendum}

\begin{figure*}[htbp]
	\centering
	\includegraphics[width=16.8 cm]{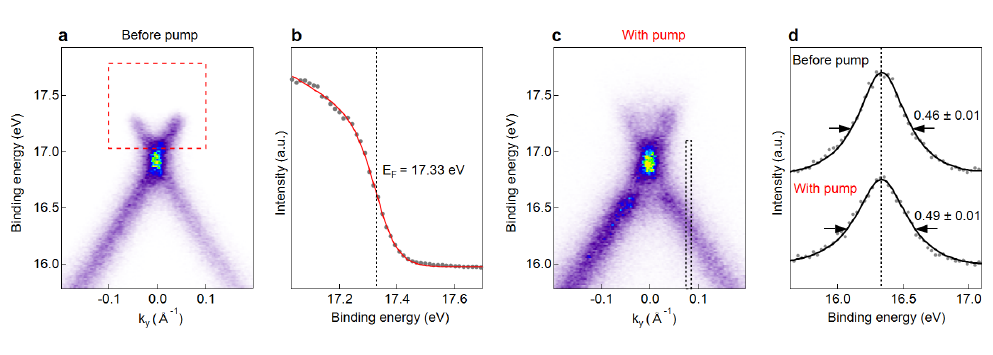}
	\caption*{\textbf{Extended Data Fig.~1 ${\mid}$ Calibration of the Fermi energy $E_F$ and negligible pump-induced band broadening at high binding energy.} 
  \textbf{a}, Dispersion image measured before pumping. 
  \textbf{b}, Extracted EDCs within the region marked by the red box and fitting using Fermi-Dirac distribution (red curve), from which $E_F$ is extracted. 
  \textbf{c}, Dispersion image measured upon pumping. 
  \textbf{d}, EDCs cut through black box at different delay times with fitting peaks appended.
	}\label{FigS1}
\end{figure*} 

\begin{figure*}[htbp]
	\centering
	\includegraphics[width=16.8 cm]{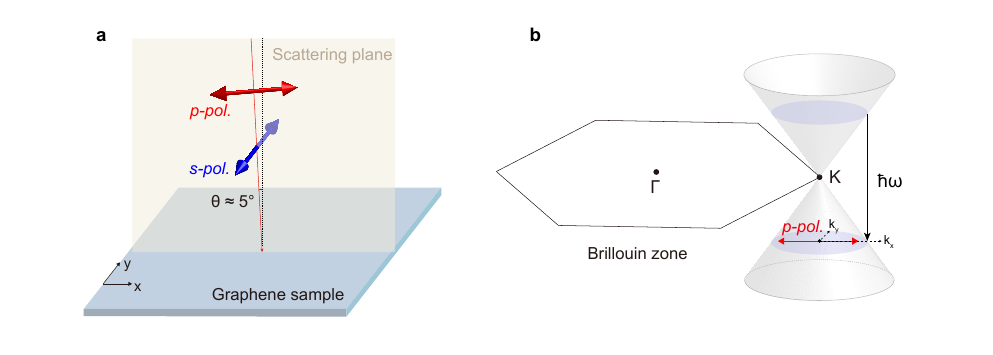}
	\caption*{\textbf{Extended Data Fig.~2 ${\mid}$ Schematic illustration of the experimental geometry.} 
  \textbf{a}, Schematic for the experimental geometry. The pump is incident on the monolayer graphene sample at near normal angle, therefore, electric fields for both \textit{p-pol.} and \textit{s-pol.} are dominantly confined within the sample plane, and there is negligible out-of-plane light field.
  \textbf{b}, Schematic illustration of the polarization with respect to the Dirac cone. The light field is along the $k_x$ ($k_y$) direction for \textit{p-pol.} (\textit{s-pol.}) pump.
	}\label{FigS1}
\end{figure*} 

\begin{figure*}[htbp]
	\centering
	\includegraphics[width=16.8 cm]{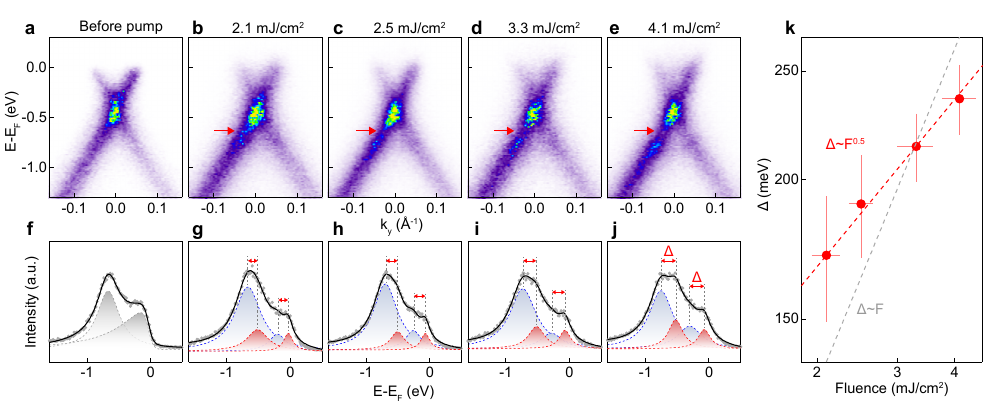}
	\caption*{\textbf{Extended Data Fig.~3 ${\mid}$ Dependence of Floquet-induced hybridization gap on the pump fluence.} 
  \textbf{a-e}, Dispersion images measured before pumping (\textbf{a}) and at $\Delta t$ = 0 fs upon driving at 490 meV with different pump fluence (\textbf{b-e}). 
  \textbf{f-j}, EDCs for data shown in (\textbf{a-e}) at momentum of resonant points.
  \textbf{k}, Extracted Floquet-induced hybridization gap as a function of the pump fluence.
	}\label{FigS2}
\end{figure*} 

\begin{figure*}[htbp]
	\centering
	\includegraphics[width=16.8 cm]{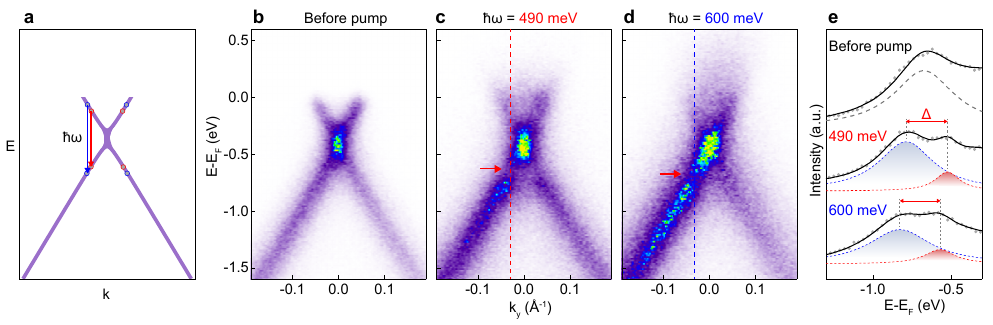}
	\caption*{\textbf{Extended Data Fig.~4 ${\mid}$ Floquet band engineering upon driving at different pump photon energies.} 
  \textbf{a}, A schematic for the resonance points upon driving with different pump photon energies.
  \textbf{b}, TrARPES dispersion images measured at $\Delta$t = -300 fs. 
  \textbf{c}, TrARPES dispersion images measured at $\Delta$t = 0 fs upon 490 meV pumping. The pump polarization is along the $k_x$ direction and the pump fluence is 4.1 mJ/cm$^2$.
  \textbf{d}, TrARPES dispersion images measured at $\Delta$t = 0 fs upon 600 meV pumping. The pump polarization is along the $k_x$ direction and the pump fluence is 6.4 mJ/cm$^2$.
  \textbf{e}, EDCs for data shown in (\textbf{b-d}) at momentum resonance point.
	}\label{FigS3}
\end{figure*} 

\begin{figure*}[htbp]
	\centering
	\includegraphics[width=16.8 cm]{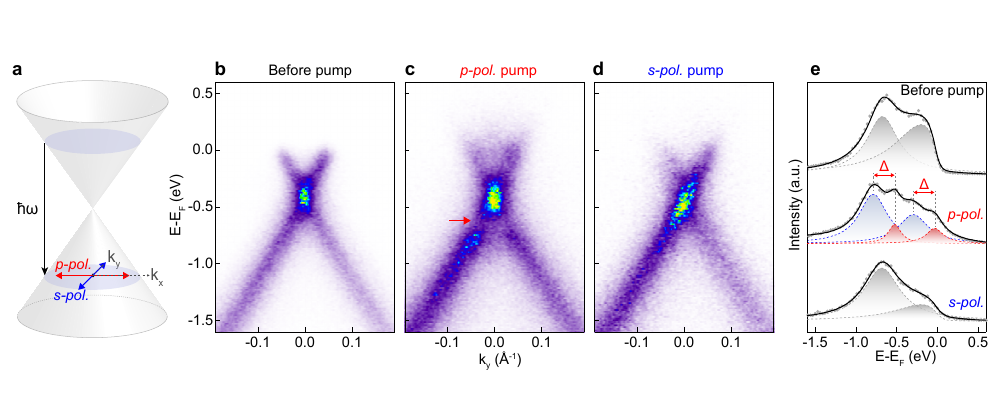}
	\caption*{\textbf{Extended Data Fig.~5 ${\mid}$ Floquet band engineering upon driving at different pump photon energies.} 
  \textbf{a}, A schematic of the experimental geometry for the \textit{p-pol.} and \textit{s-pol.} pumps.
  \textbf{b-d}, TrARPES dispersion images measured at $\Delta$t = -300 fs (\textbf{b}) and $\Delta$t = 0 fs with \textit{p-pol.}  pump (\textbf{c}) and \textit{s-pol.}  pump (\textbf{d}). The pump photon energy is 490 meV and the pump fluence is 4.1 mJ/cm$^2$.
  \textbf{e}, EDCs for data shown in (\textbf{b-d}) at momentum resonance points with fitting peaks appended.
	}\label{FigS5}
\end{figure*} 

\begin{figure*}[htbp]
	\centering
	\includegraphics[width=14.8 cm]{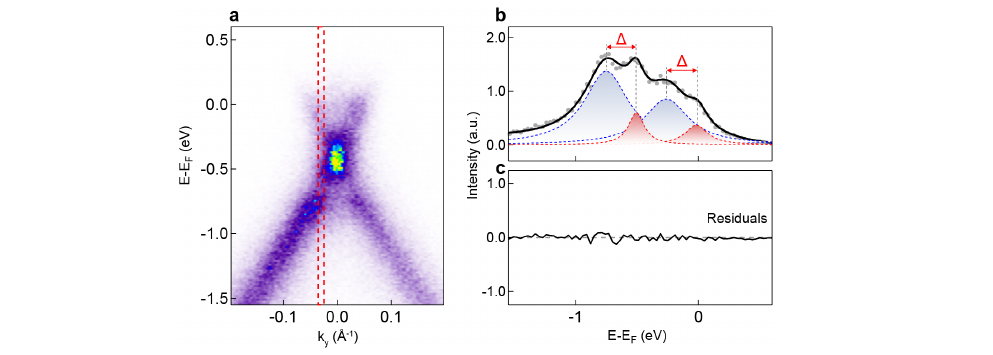}
	\caption*{\textbf{Extended Data Fig.~6 ${\mid}$ Fitting residuals to show the quality of the fitting.}
  \textbf{a}, Dispersion image at $\Delta t = 0$, with the momentum integration window marked by lines. 
  \textbf{b}, EDCs obtained by integrating over the indicated momentum range.  \textbf{c}, Residuals from the EDC fitting, shown on the same scale as (\textbf{b}).
	}\label{FigS6}
\end{figure*} 

\begin{figure*}[htbp]
	\centering
	\includegraphics[width=16.8 cm]{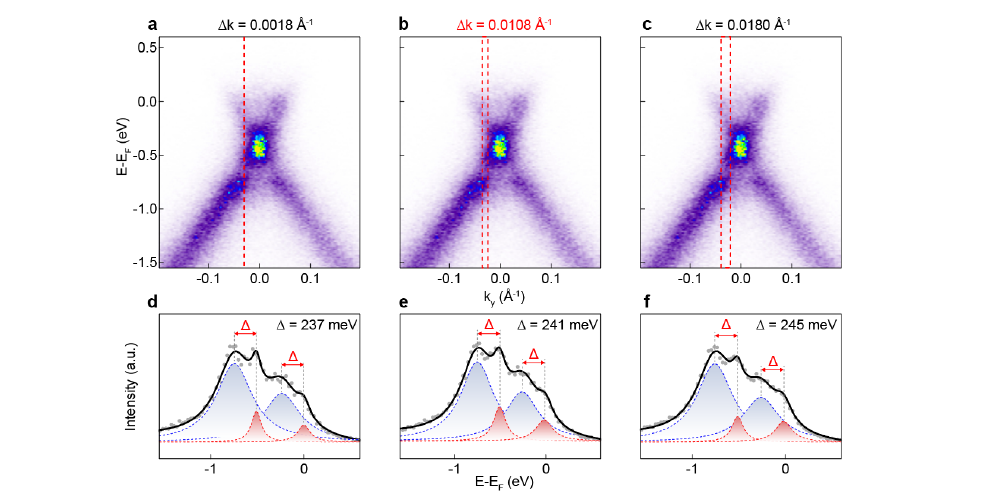}
	\caption*{\textbf{Extended Data Fig.~7 ${\mid}$ Variation of the extracted gap by using different choices of momentum integration windows.}
  \textbf{a-c}, Dispersion images, where dashed boxes indicate the momentum integration windows used for EDCs shown in (\textbf{d-f}). The values are also labeled on the top of each panel. 
  \textbf{d-f}, Extracted gap values obtained from EDC fittings.
	}\label{FigS7}
\end{figure*} 

\begin{figure*}[htbp]
	\centering
	\includegraphics[width=16.8 cm]{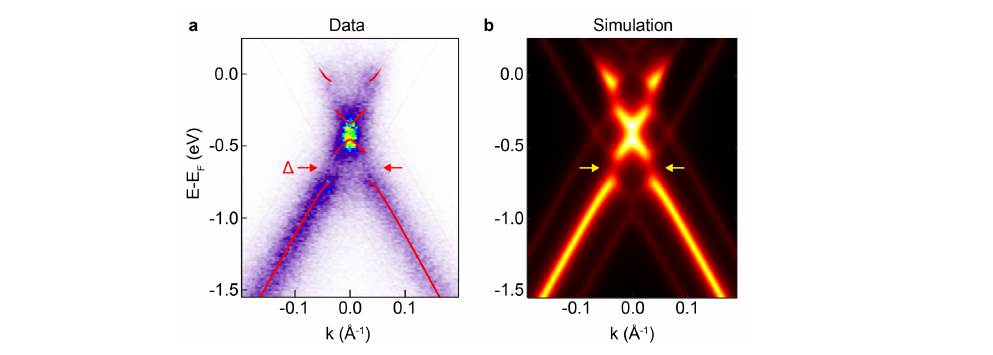}
	\caption*{\textbf{Extended Data Fig.~8 ${\mid}$ TrARPES dispersion image and simulation.}
  \textbf{a}, Dispersion image at $\Delta t = 0$ with simulated dispersion over plotted. \textbf{b}, Simulated TrARPES spectrum upon pumping.
	}\label{FigS8}
\end{figure*}

\begin{table*}
    \centering
    \renewcommand{\arraystretch}{1.4}
    \setlength{\tabcolsep}{10pt}
    \resizebox{\textwidth}{!}{
    \begin{tabular}{|c|c|c|c|c|c|}
        \hline
          & \makecell{S. Aeschlimann \\\textit{et al.} (Ref.\cite{GierzNanoLett2021})}
 & \makecell{M. Merboldt \\\textit{et al.} (Ref.\cite{stefan2025graphene})}
 & \makecell{D. Choi \textit{et al.} \\(Ref.\cite{gedik2025graphene})}
 & \multicolumn{2}{c|}{\textbf{This work}}\\
        \hline
        \textbf{Pump photon energy} $\hbar\omega$ (meV)          
		& 280           &  650          & 246           & \textbf{490}       & \textbf{600}    \\
        \hline	
        Pump fluence $F$ (mJ/cm\textsuperscript{2}) 
        & 2          &  1.23      &   0.0254        & 4.1       &6.0\\
        \hline
        \textbf{Pump electric field in the vacuum} $E$ (V/m)
        & 2.4 $\times$ 10$^8$          &   3 $\times$ 10$^8$         &  0.27 $\times$ 10$^8$          & \textbf{4.8 $\times$ 10$^8$ }      &\textbf{6.3 $\times$ 10$^8$ }  \\   
        \hline																	  \textbf{In-plane electric field inside the sample} $E_i$ (V/m) 
	    & /          &   /         &  0.082 $\times$ 10$^8$          & \textbf{1.8 $\times$ 10$^8$ }  & \textbf{2.4 $\times$ 10$^8$ } 											\\  
        \hline
        \textcolor{red}{\textbf{Floquet parameter }$\beta$                           } & /             & 0.108         & 0.085         & \textcolor{red}{\textbf{0.49}}   & \textcolor{red}{\textbf{0.44}}      \\
        \hline
        \textcolor{red}{\textbf{Floquet-induced hybridization gap }$\Delta$ (meV)}                           & /             &70 (cal.)     &  21 (cal.)     & \textcolor{red}{\textbf{240} (cal.)}  & \textcolor{red}{\textbf{263} (cal.)} \\
        \hline
        Energy resolution $\Delta E$ (meV)                   & 150           & 155          & 53             & 71       & 71    \\
        \hline
        \textcolor{red}{\textbf{Ratio between $\Delta$ and $\Delta E$} ($\frac{\Delta}{\Delta E})$                                            } & /             & 0.45        &   0.40          & \textcolor{red}{\textbf{3.38}}  & \textcolor{red}{\textbf{3.70}} \\
        \hline
        Incidence angle $\alpha$ ($^\circ$)      & /             &  22          &  8.7        & 5     & 5       \\
        \hline
        Pump pulse duration $\Delta\tau$ (fs)       & 270           &   100         & 247           & 135     & 113	    \\
        \hline
        Probe pulse duration $\Delta\tau_{pr}$ (fs)       & 120           & 20           &  247          & 66  & 66            \\
        \hline
    \end{tabular}
    } 
	\caption*{\textbf{Extended Data Table 1 ${\mid}$ Comparison of our experimental parameters with those in the literature.}  All parameters in the literature were adapted from Ref.\cite{GierzNanoLett2021,gedik2025graphene,stefan2025graphene}. The Floquet parameter $\beta$ is calculated from the in-plane electric field inside the sample $E_i$ and pump photon energy $\hbar\omega$ by $\beta=\dfrac{e v_f E_i}{\hbar \omega^2}$, except for Ref.\cite{stefan2025graphene} where $\beta$ is deduced from $\hbar\omega$ and $\Delta$ by $\beta=\frac{\Delta}{\hbar\omega}$. }
\end{table*}

\end{document}